\documentclass[twocolumn,prl,showpacs]{revtex4}
\usepackage{graphicx}
\usepackage{rotating}
\usepackage{amsmath}
\usepackage{amsfonts}
\usepackage{amssymb}
\usepackage{enumerate}
\usepackage{longtable}
\usepackage{dcolumn}
\usepackage{bm}
\usepackage{epsfig}
\usepackage{subfigure}
\begin{document}

\title{Preformed excitonic liquid route to a charge density 
wave in 2H-TaSe$_2$ }
\author{A. Taraphder$^{1,2}$, S. Koley$^{1}$, N. S. Vidhyadhiraja$^{3}$
and M. S. Laad$^{4}$}
\affiliation{$^{1}$Department of Physics and Centre for Theoretical studies,\\
Indian Institute of Technology, Kharagpur 721302 India \\
$^{2}$Max-Planck-Institut f\"ur Physik Komplexer Systeme,
 N\"othnitzer Str. 38, 01187 Dresden, Germany\\
$^{3}$Jawaharlal Nehru Centre for Advanced Scientific Research, Bangalore
560064 India\\
$^{4}$ Inst. f\"ur Theor. Physik. 1B, Rheinische-Westfalische
Technische Universit\"at D 52056 Aachen, Germany }


\begin{abstract}
Recent experiments on 2H-TaSe$_2$ contradict the long-held view of the 
charge density wave arising from a nested band structure. An intrinsically strong 
coupling view, involving a charge density wave state arising as a Bose 
condensation of preformed excitons emerges as an attractive, albeit 
scantily investigated alternative. Using local density approximation 
plus multiorbital dynamic mean-field theory, we show 
that this scenario agrees with a variety of normal state data for 2H-TaSe$_{2}$.
Based thereupon, the ordered states in a subset of dichalcogenides should be 
viewed as instabilities of a correlated, preformed excitonic liquid.  
\end{abstract}
\maketitle

Discovery of superconductivity in layered transition-metal dichalcogenides 
(TMD) on doping~\cite{cava} and under pressure~\cite{sipos} recently have 
rekindled interest~\cite{wilson1,mcwh,aebi} in them.     
Presence of valence and conduction bands with different orbital 
symmetries near the Fermi energy ($E_{F}$) and sizable electron-electron 
interactions on a triangular lattice produce fine tunability amongst 
competing broken symmetry states. Strong collective fluctuations in the 
nearly degenerate manifold of states may give rise to novel 
phases at low temperature ($T$) in response to minute changes in external 
stimuli. These features also lead to 
poorly understood bad metallic~\cite{at08} and non-Fermi liquid (nFL) 
behavior~\cite{si} in putative normal states. 

The ubiquitous charge density wave (CDW) instabilities in TMD have 
long been rationalized as consequences of Fermi surface (FS) 
nesting~\cite{wilson1,scot}. Upon 
closer experimental scrutiny recently~\cite{liu,dard,straub,ruz}, however, 
this mechanism appears unlikely in 2H-TaSe$_2$, which shows
incommensurate (ICDW, T$_{ic} = 122$K) and commensurate CDW (CCDW, T$_{cc}=90$K) 
transitions~\cite{wilson1,mcwh}. Lack of correlation between the charge  
susceptibility at the  nesting vector and the CDW wavevector in ARPES~\cite{boris1} 
and near absence of changes in band dispersion across $T_{ic},\, 
T_{cc}$~\cite{liu,valla} are difficult to reconcile with FS nesting.  

Additional evidence comes from~\cite{bark} recent 
transport data in 2H-TaSe$_2$ showing no pronounced anomaly at 
$T_{ic,cc}$~\cite{dord,vesc,ruz}. Apart from a perceptible change of slope  
at T$_{ic}$, the in-plane resistivity $\rho_{ab}$ is almost linear~\cite{ruz} 
from T$_{ic}$ to about 400K with a slight change of slope around 300K 
($\rho_{c}$ nearly follows $\rho_{ab}$, though 25-50 times larger). 
The small transport mean-free paths, 
$l_{c} < a$ (the lattice spacing) and $l_{ab}\simeq 5a$ \cite{ruz}, indicate a  
typical {\it bad metal}. Remarkably, despite the absence of well-defined 
Fermi liquid (FL) quasiparticles in transport, ARPES spectra fit nicely  
with a {\it local self-energy} $\Sigma(\omega)$   
without invoking any phonon coupling~\cite{valla}. Transport 
scattering rates~\cite{vesc} $\tau_{tr}$ and quasiparticle life 
time $\tau_{qp}$ from ARPES 
show similar variation~\cite{valla} with $T$, confirming dominant 
local correlations. This is consistent with observed 
agreement between LDA bands and ARPES dispersion, implying a negligible 
$k$-dependence of $\Sigma(\omega)$, but {\it not} weak electronic 
correlations. Optical conductivity, $\sigma(\omega)$ reveals the formation 
of a pseudogap around 300K, progressive narrowing of the ``Drude'' peak in the 
far infrared region for $T < T_{cc}$, and sizable spectral weight transfer (SWT) 
with $T$.  These are generic
fingerprints~\cite{valla,dord} of sizable local electronic correlations in the 
normal state.  

Thus, extant data reveal an incoherent bad metal relieving its entropy at lower 
$T$ by transforming either to 
an unconventional CDW (UCDW) or an unconventional superconducting (USC) state. 
The issue is thus: {\sl What causes normal state 
incoherence, and how do UCDW/USC states arise from such a high $T$ state?}  
We show that these features in 2H-TaSe$_{2}$ can be 
semi-quantitatively understood within a new, intrinsically 
strong coupling picture where UCDW/USC states are 
viewed as instabilities of an incoherent, preformed excitonic liquid.
This alternative view remains, to our knowledge, largely unexplored. 

LAPW band structure~\cite{smith} of $2H$-TaSe$_2$ shows a negative 
indirect band gap with six hole pockets in FS and a strong 
Ta $d_{z^{2}}$ character in the two bands crossing $E_F$.  LCAO results 
with Ta 5d and Se 4p orbitals gives the two bands closest to $E_{F}$, the Se 
($p_{z}$ predominantly) and the Ta (predominantly $d_{z^{2}}$) bands as 
well as the FS in excellent accord with 
LAPW. A sizable $d_{z^{2}}$-$p_{z}$ mixing ($t_{ab}$) mixes the small 
number of electrons and holes. Although bare Coulomb interactions are 
not large ($<1.0$~eV), given small carrier density, even a moderate 
{\it interband} $U_{ab}$ facilitates exciton formation at high $T$~\cite{rice}. 
Observed normal state incoherence constrains one to adequately treat
local correlation effects in 
multiorbital (MO) Hubbard-like model for 2H-TaSe$_2$~\cite{sk_10}, which we 
solve using DMFT~\cite{kotliar} with the LCAO density-of-states (DOS) as 
input. The Hamiltonian is

\begin{eqnarray*}
H &=& \sum_{{\bf k} a b \sigma}(t_{{\bf k}ab}
+\epsilon_{a}^{0}\delta_{ab})c_{{\bf k}a\sigma}^{\dag}c_{{\bf k}b\sigma} 
+ U\sum_{ia}n_{ia\uparrow}n_{ia\downarrow} \\ 
&+& U_{ab}\sum_{i a \ne b} n_{ia}n_{ib} 
\end{eqnarray*}
\noindent where $a,b$ denote the LDA conduction ($d$) and valence ($p$) 
bands with dispersions $t_{aa}, \, t_{bb}$. $t_{ab}\, (a\ne b)$ is the mixing 
of the two bands and $U,\, U_{ab}$ are intra and interorbital local Coulomb 
interactions. From LCAO, the orbital character of the $d$ band crossing FL 
is predominantly d$_{z^2}$ at $\Gamma$ point with admixture of d$_{xy}$/d$_{x2-y2}$ 
at K point, consistent with LDA~\cite{smith}.  
We solve $H$ by LCAO + MO-DMFT using iterated perturbation 
theory (IPT), used successfully for transition metal 
oxides~\cite{laad}. Local dynamical correlations (in DMFT) modify the LCAO 
bands in two major ways: First, the intra and interorbital Hartree terms 
renormalize the relative band positions. Simultaneously, dynamical 
correlations cause SWT across large energy, missed 
by static mean-field theory, which, {\it cannot}, therefore, access 
incoherent states.  

\begin{figure}


{\includegraphics[angle=270,width=0.7\columnwidth]{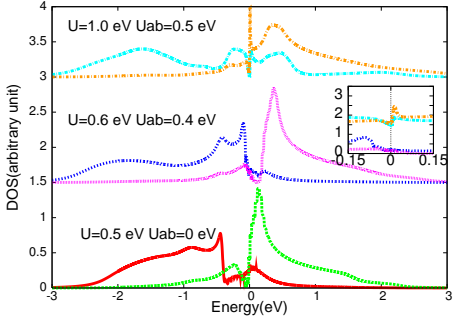}
}

\caption{(Color online)
Evolution of the many-particle DOS with different $U$ and $U_{ab}$
for $t_{ab}=0.4$~eV. Inset shows ``gap'' formation at E$_{F}$ with finite $U_{ab}$. 
The $p$ \,($d$) band DOS are mostly below (above) E$_F$.}   
\label{fig1}
\end{figure}

In Fig.1, we show the LCAO+DMFT results for a combination of  
$U=U_{aa}\simeq U_{bb}$ and $U_{ab}$ for $t_{ab}=0.4$~eV (results are 
essentially insensitive to reasonable variation of these). Given the small ($d$) 
electron and ($p$) 
hole densities, neglecting $t_{ab}$ merely shifts the Se $p$ band totally 
below $E_{F}$ without significant modification of the spectra. Once the $p$ band 
is pushed below $E_{F}$, the excitonic average, $\langle c_{ia\sigma}^{\dag} 
c_{ib\sigma}\rangle$ vanishes identically (Elitzur's theorem). 
Interestingly, the van Hove feature of the
Ta $d$ band remains pinned to $E_{F}$: by itself, this could generate a 
nesting-induced CDW solely involving the Ta $d$  band at low $T$. For such 
a weak modification, the normal state would be a moderately 
correlated FL, in stark conflict with experimental data on 2H-TaSe$_{2}$, 
though it could conceivably be relevant for other cases~\cite{at08}. 

\begin{figure}
{(a)}
{\includegraphics[angle=270,width=0.43\columnwidth]{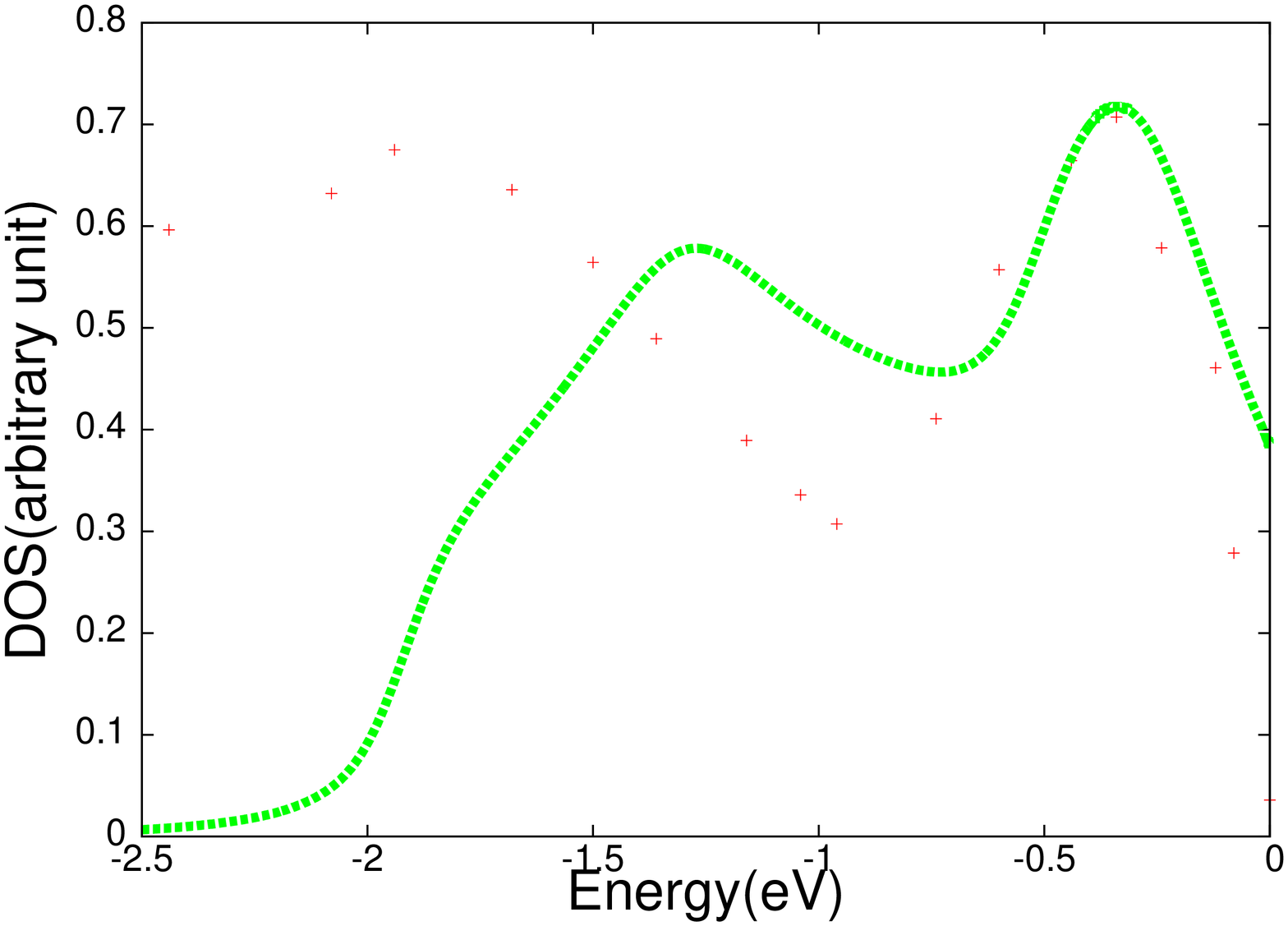}}
{(b)}
{\includegraphics[angle=270,width=0.43\columnwidth]{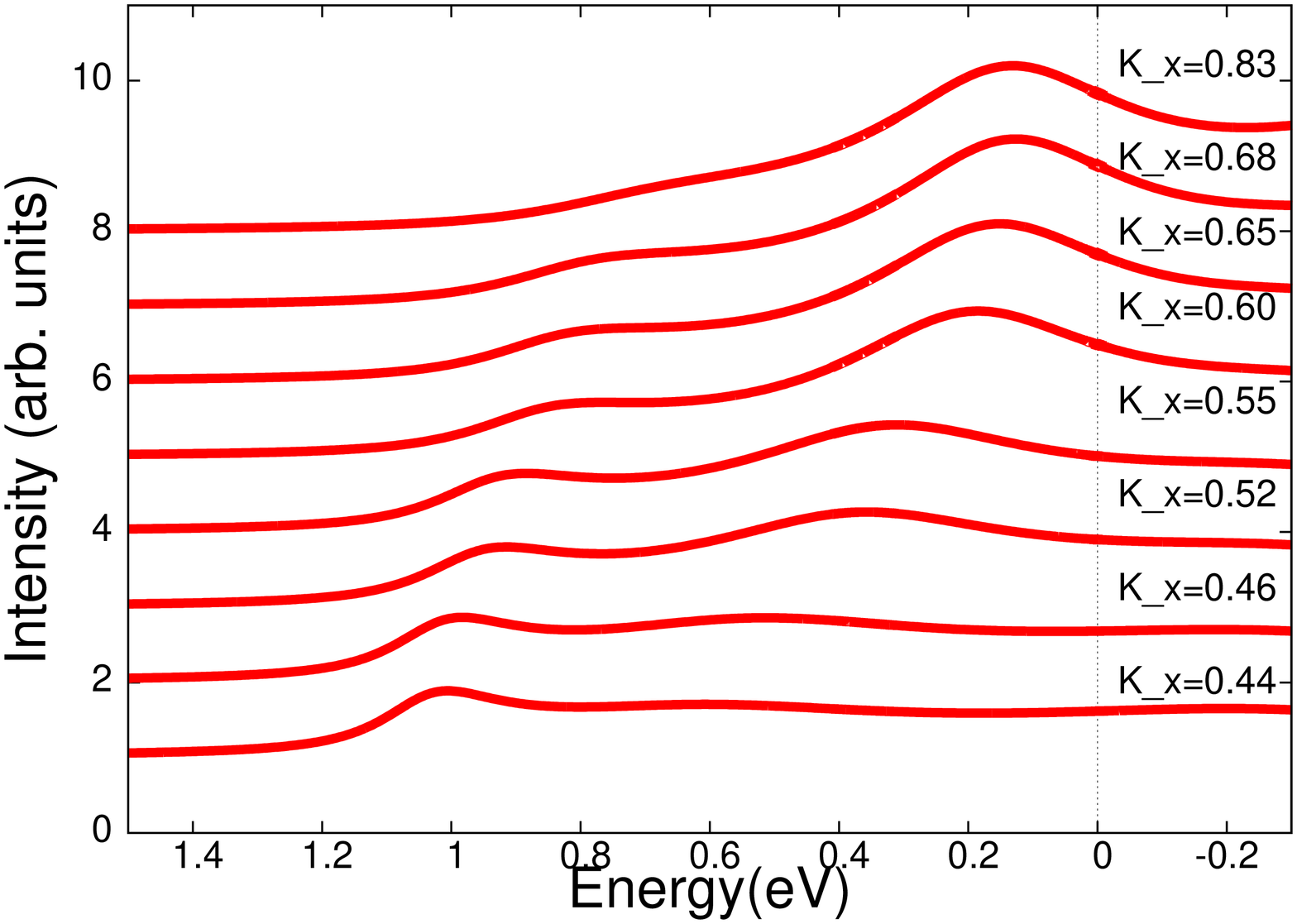}}
\caption{(Color online) 
Comparison of DMFT DOS (solid line) at $T=300$K
with photoemission data~\cite{okuda} (dotted curve) clearly
showing good quantitative agreement. (b) Evolution of the
{\bf k}-dependent spectral
function in a direction similar to Liu et al.~\cite{liu}.
}
\label{fig2}
\end{figure}

Inclusion of $t_{ab}$ (=0.4~eV) drastically modifies above results. A clear
low-energy pseudogap instead of a quasiparticle pole, along with high-energy 
Hubbard bands, is discernible in the DOS (Fig.1). 
This excellently describes 
photoemission (PES) data
\cite{okuda} up to -1.5~eV (Fig.2a). In Fig.2b, we show the theoretical ARPES
lineshapes, clearly reflecting (dynamically) renormalized 
Ta $d$ states dispersing 
through $E_{F}$. Though this ${\bf k}$-dependent feature is captured
by LDA~\cite{liu,valla}, the experimental ARPES linewidths 
are too broad, reflecting the incoherent metal features. 
The latter, coming from strong dynamical 
correlations, is only captured by DMFT. Moreover, we predict
that ARPES measurements up to high binding energy will reveal the lower
Hubbard band around -1.5 to -2.0~eV.

ARPES data also show a gradual build-up of excitonic correlations as  
the pseudogap deepens~\cite{boris1} and the low-energy peak in PES shifts 
to higher energy, accompanied by a $T$-induced SWT.  
Our DMFT results (Fig.3) track ARPES data in all aspects, 
including the sizable SWT and details of the lineshape. 
Revealingly, setting $t_{ab}=0$ (inset Fig.3) disagrees with data: the 
valence band peak lies {\it above} $E_{F}$ at high $T$, and 
no pseudogap is discernible at lower $T$. 
Finally, Im$\Sigma(\omega)$ also shows a drastic reduction of incoherence 
with progressive stabilization of exciton-induced 
pseudogap as $T$ reduces.  

Normal state transport in 2H-TaSe$_{2}$ also finds comprehensive explication 
within our theory. Fig.4a, shows our DMFT results for the optical 
conductivity, $\sigma(\omega,T)$. LDA+DMFT calculations~\cite{kot-vo2}  
without the vertex corrections for multiorbital cases give a quantitatively 
accurate estimate of $\sigma (\omega)$. 
Even though finite, we expect a small contribution from vertex corrections, and 
neglect it. Quite remarkably, the $\omega$ {\it and} $T$ dependence 
of $\sigma(\omega,T)$ are faithfully captured by DMFT. 
Although correlated FL behavior is never found, gradual 
build-up of excitonic coherence in tandem with reduced 
incoherent scattering at lower $T$ is clearly reflected
in $\sigma(\omega,T)$. At low $T$, a weak shoulder-like
feature around $0.4$~eV demarcates the scale below which enhanced coherence
sets in - the same scale at which additional gap-like features
appear in the DOS (Fig.3), establishing that 
increasing low-energy coherence in $\sigma(\omega,T)$ at low $T$ reflects
that of the preformed excitons. At higher $T$, this shoulder in
$\sigma(\omega,T)$ rapidly disappears, signifying a rapid 
crossover to incoherent excitonic regime.  

\begin{figure}
\centering
\includegraphics[angle=270,width=0.49\columnwidth]{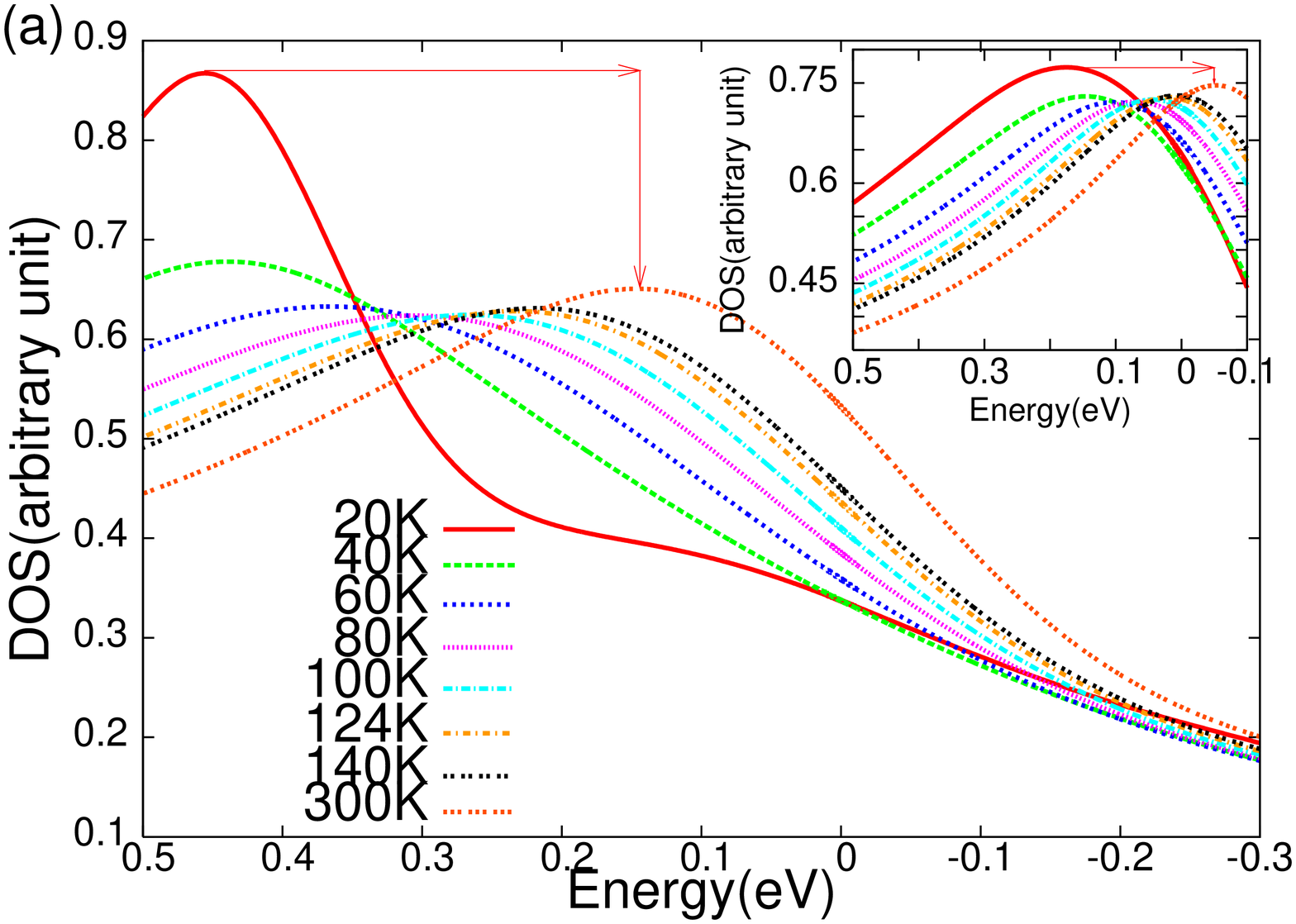}
\includegraphics[angle=270,width=0.49\columnwidth]{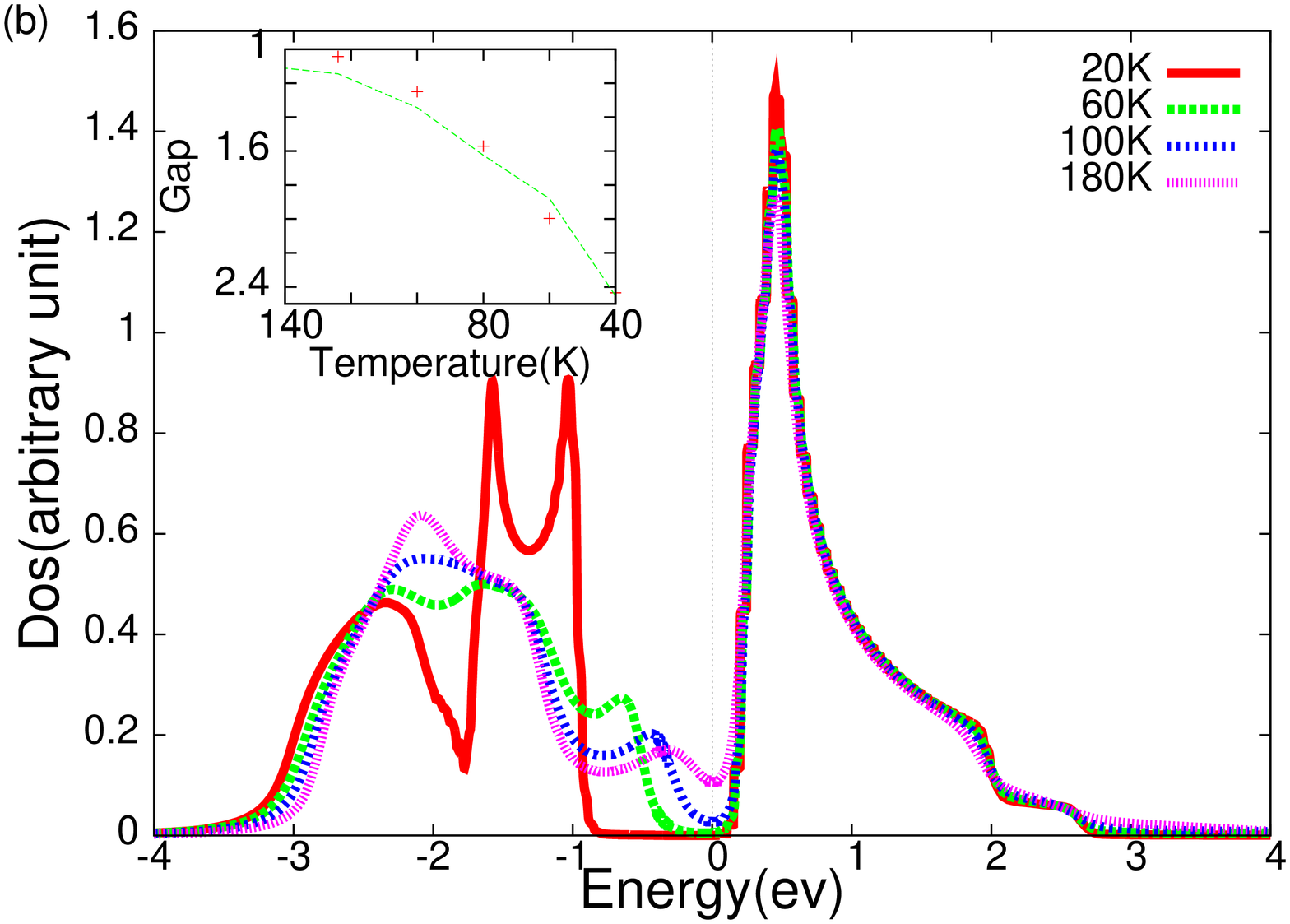}
\caption{(Color online) Evolution of the spectral function with $T$
and formation of the gap (see text) (a) with ($t_{ab}$=0.4) and without
($t_{ab}$=0.0, inset) preformed excitons, (b) with CDW order. Inset shows a fit
to the normalized ``band gap" (``+" symbols \cite{boris1}) from our calculation (line).
}
\label{fig3}
\end{figure}

Fig.4b shows our DMFT results for the $T$-dependent $dc$ resistivity. In 
accord with experimental data, no FL regime is found: instead, 
$\rho(T)$ shows a
broad bump around 100~K, below which enhanced metallicity
is recovered ($\rho(T)$ still varies nearly linearly with $T$). When 
$t_{ab}=0$, qualitatively similar behavior at 
high $T>300$~K, smoothly evolves into $\rho(T)\propto  
T^{2}$ at low $T$ as in a correlated FL with reduced bump, in stark 
conflict with data~\cite{vesc,dord}. Thus, strong scattering off incoherent 
preformed excitons wipes out FL coherence. It also provides a rationale for the 
insensitivity of transport to the onset of CDW order: if excitonic 
correlations already establish themselves
at high $T$, most of the band FS already gets modified to reflect 
the preformed, incoherent excitons. Additional FS changes at the CDW 
transition are then small enough that transport will not  
see the onset of CDW. A large $2\Delta/k_{B}T_{cc}>10$ ratio found~\cite{dard} 
in 2H-TaSe$_{2}$ fully supports this view: 
this implies~\cite{mcmillan} that (i) strong scattering dominates the 
normal state, and (ii) transport is less sensitive to onset of 
LRO, but will show clear precursor features in the normal state of 
our DMFT. Additionally, the $T$-variation of the carrier scattering rates 
(inset Fig.4b), is consistent with the reported fit~\cite{valla} to the high-$T$ 
transport ($T>120$~K). 
Finally, given the in-plane normal state incoherence (with an exciton-induced
pseudogap), the out-of-plane responses will show even more drastic signatures 
of incoherence, as indeed observed~\cite{dord}.
 
\begin{figure}
{(a)}
{\includegraphics[angle=270,width=0.7\columnwidth]{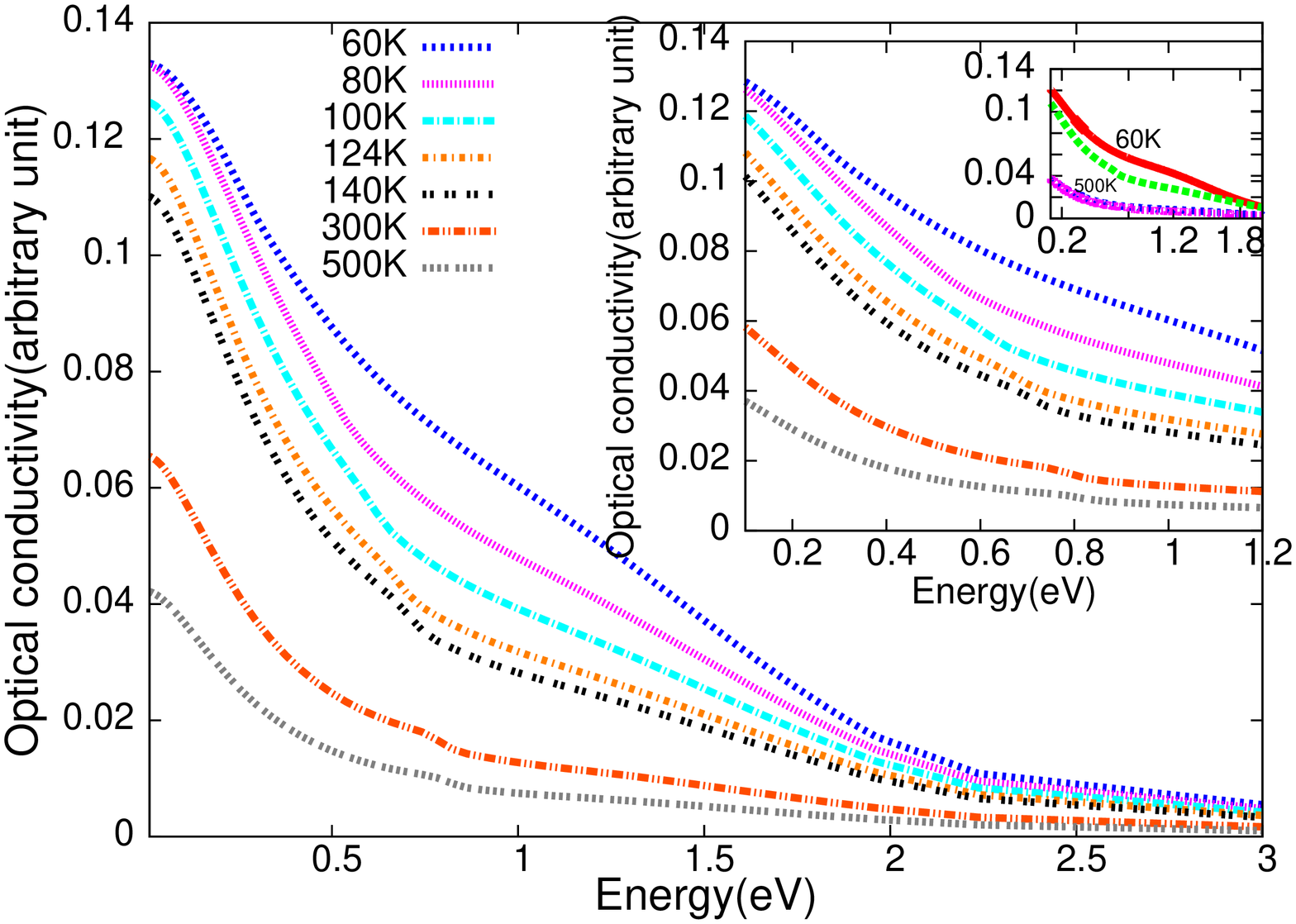}
}

{(b)}
{\includegraphics[angle=270,width=0.65\columnwidth]{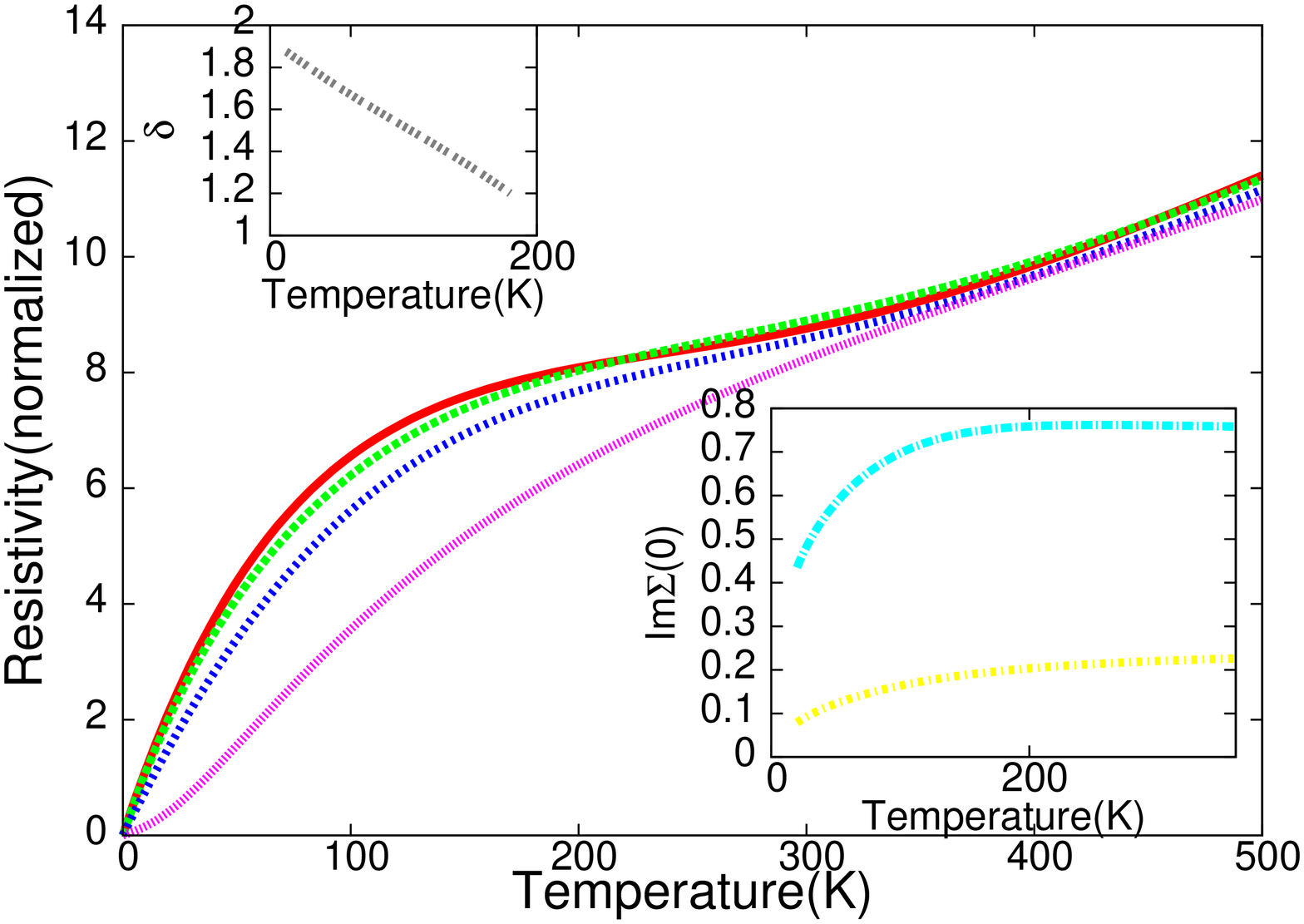}
}
\caption{(Color online) (a) Calculated $\sigma(\omega)$ ($t_{ab}=0.4$) at 
various $T$. The same at low energy (inset) and 
with and without excitonic effects at 60K and 500K (small inset).
(b) $\rho_{dc}$ for $t_{ab}$=0.4: with CDW, for $t_{ab}$ =0.3 and $t_{ab}$=0.0  
(upper to lower). Right inset: $Im\Sigma(\omega=0)$ versus $T$ for the two 
bands, and the left for CDW $T^z$.   
}  
\label{fig4} 
\end{figure} 

If the preformed excitonic liquid idea is to hold, changes in spectral and 
transport data should bear a one-to-one correlation 
with $T$ and $\omega$-dependent evolution of the excitonic spectral 
function, $\rho_{ab}(\omega)=(-1/\pi) $Im$G_{ab}(\omega)$ and the local 
excitonic amplitude, $\langle (c_{ia\sigma}^{\dag}c_{ib\sigma}+h.c)\rangle=(-1/\pi)
\int_{-\infty}^{\infty}d\omega Im G_{ab}(\omega)$. 
The strong $T$-dependence of $\rho_{ab}(\omega)$ within DMFT is obvious 
(Fig.5, inset): at 
high $T$, the broad, asymmetric shape is a manifestation of the incoherent 
excitonic fluid, while the low-energy pseudogap and large SWT  
with decreasing $T$ signal a build-up of incipient
excitonic coherence. This is seen in the steep increase of the excitonic 
amplitude below 100~K (Fig.5), and correlates with the broad bump in 
$\rho(T)$ in Fig.4b, testifying the strong, dynamic excitonic correlations in 
2H-TaSe$_2$.  

\begin{figure}
{\includegraphics[angle=270,width=0.7\columnwidth]{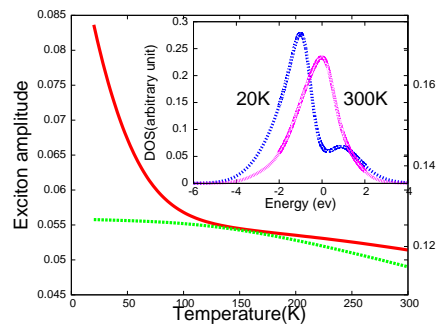}}
\caption{(Color online) Variation of excitonic amplitude with $T$ in the 
presence (dotted, green: right scale) and absence (line, red: left scale) of 
CDW order. Local excitonic spectral function (inset, see text) at 20K and 300K.}

\label{fig5}
\end{figure}
A natural question, therefore, is how do we understand the CDW/SC
found in TMD at low $T$?
If these arise from a high-$T$ incoherent metal, as proposed here,
they cannot be viewed as instabilities of an FL, as coherent FL
quasiparticles are unstable at the outset. In analogy with coupled luttinger
liquids, in a regime where the one-particle mixing term
($t_{ab}$) is irrelevant, two-particle coherence at
second order~\cite{giam} in $t_{ab}$, in p-h (CDW) or p-p (SC) channels
will arise from intersite and interband pair-hopping terms via
$H'\simeq t_{ab}^{2}\sum_{<i,j>}\chi_{ij}^{ab}(\omega)(c_{ia\sigma}^{\dag}
c_{ib\sigma}c_{jb\sigma'}^{\dag}c_{ja\sigma'}+h.c)$, where
$\chi_{ij}^{ab}(\omega)$ is the dressed excitonic susceptibility, estimated from
the {\it normal} state DMFT results.  Instabilities to UCDW/USC states
occur upon a Hartree-Fock (HF) decoupling of
$H'$ in p-h and p-p channels, a procedure {\it literally} exact in
DMFT, since $H'\simeq O(1/D)$.
Starting with the new Hamiltonian, $H=H_{n}+H_{res}^{HF}$, where
$ H_{n} = \sum_{k,\nu}(\epsilon_{k,\nu}+\Sigma_{\nu}(\omega)-E_{\nu})c_{k,\nu}^{\dagger}
c_{k,\nu}+\sum_{a\ne b,(k)}t_{ab}( c_{k,a}^{\dagger}c_{k,b}+h.c.)$,
with $\nu=a,b$ and
$H_{res}=-g\sum_{\langle i,j\rangle,a,b}(\langle n_{i,a}\rangle
n_{j,b}+\langle n_{j,b}\rangle n_{i,a}-\langle
c_{i,a}^{\dagger}c_{j,b}^{\dagger}\rangle c_{j,b}c_{i,a}+ h.c.)$, we
study the CDW phase (with parametrized but realistic $g=0.35$ \cite{mcmillan}). 
Since $T^{z}=\frac{1}{2}(n_{a}-n_{b}),\,\,
T^{+}= c_{a}^{\dagger}c_{b}\,\, {\rm and} \,\, T^{-}=
c_{b}^{\dagger}c_{a}$, onset of CDW ($T^{z}$) order results
in reduction of excitonic
($T^{+}, \, T^{-}$) liquid fluctuations, resulting in increase in
$\langle c_{a}^{\dag}c_{b}+h.c.\rangle$, as indeed seen
in Fig.5. The consequent suppression of QP
scattering rate below $T_{cdw}$ and resulting reduction in resistivity
(shown in Fig.4b),
as seen experimentally, fully corroborate
our assertion that CDW is a ``coherence-restoring'' transition.  Onset
of CDW order stabilises the ``gap'' in the normal
state DOS (Fig.3), again in nice accord with ARPES.  Finally, the small
increase in $\langle T^{z}\rangle$ around $100$~K (Fig.5)
reflects CDW order arising from a preformed excitonic state, and
qualitatively similar behavior is found in ARPES studies on 1T-TiSe$_{2}$~\cite{monney}.

Thus, the UCDW ordered state is now interpretable as a bose-condensed
phase of excitons.
Indeed, very good agreements with variety of normal state features strongly
support the preformed excitonic
view presented here, at least for 2H-TaSe$_{2}$. Since SC in many other TMDs
arises from (nearly) incoherent ``normal'' states on the border of CDW order,
the present scenario, extended to USC order, should have generic
applicability to
these cases. Such an excitonic CDW will, in reality, involve phonons as well.
However, lack of any signature of carrier-lattice coupling in ARPES suggests
that the CDW is dominantly electronically driven in 2H-TaSe$_{2}$.
Thus, our picture is not in conflict with the exciton-plus-phonon
idea. Theoretically, integrating out the phonons from terms like
$g\sum_{i,\sigma}a_{i\sigma}^{\dag}b_{i\sigma}(A_{i}+A_{i}^{\dag})$ \cite{monney}
describing exciton-phonon coupling (with $A_{1g}$ symmetry relevant to
TMD) yields
an additional contribution
$-(g^{2}/\Omega)\sum_{<i,j>}a_{i\sigma}^{\dag}b_{i\sigma}b_{j\sigma'}^{\dag}a_{j\sigma'}$
to $H_{res}$ (here, $\Omega$ is the $A_{1}$-optical phonon energy),
and thus only renormalises the {\it effective}
two-body potential
in $H_{res}$ leading to the CDW instability above.
Coupled with the absence of distinctive electron-lattice coupling
features (e.g, kinks near the relevant phonon energies) in ARPES, our work
strongly supports a {\it primary} role for preformed excitons in the
emergence of CDW order. Finally, SC at much lower $T_{sc}\simeq
200$~mK has been reported
in literature~\cite{maaren}.  This can be studied using the pairing
term in $H_{res}$.  It may well turn out that
$T_{sc}$ can be enhanced by pressure, but this demands more
experimental and theoretical work.
\vspace{2mm}

SK acknowledges CSIR (India) for a fellowship. We thank H. Beck, 
P. B. Littlewood, S. Saxena and C. M. Varma for helpful discussions.

\end{document}